\documentclass[twocolumn,aps,floats,floatfix,groupedaddress]{revtex4-1}
\usepackage{graphicx}
\usepackage{dcolumn}
\usepackage{amsmath}
\usepackage{amsfonts}
\usepackage{amssymb}
\usepackage{epsfig,float,afterpage}
\usepackage{natbib}

\begin{document}
\title{Spin noise spectroscopy beyond thermal equilibrium and linear response}
\author{P. Glasenapp$^1$, Luyi Yang$^2$, D. Roy$^{3}$, D. G. Rickel$^2$, A. Greilich$^1$, M. Bayer$^1$, N. A. Sinitsyn$^3$, S. A. Crooker$^2$}

\affiliation{$^1$Experimentelle Physik 2, Technische Universit\"at Dortmund, D-44221 Dortmund, Germany}

\affiliation{$^2$National High Magnetic Field Lab, Los Alamos National Laboratory, Los Alamos, New Mexico 87545, USA}

\affiliation{$^3$Theoretical Division, Los Alamos National Laboratory, Los Alamos, New Mexico 87545, USA}


\begin{abstract}
Per the fluctuation-dissipation theorem, the information obtained from spin fluctuation studies in thermal equilibrium is necessarily constrained by the system's linear response functions. However, by including weak radiofrequency magnetic fields, we demonstrate that intrinsic and random spin fluctuations even in strictly unpolarized ensembles \emph{can} reveal underlying patterns of correlation and coupling beyond linear response, and can be used to study non-equilibrium and even multiphoton coherent spin phenomena. We demonstrate this capability in a classical vapor of $^{41}$K alkali atoms, where spin fluctuations alone directly reveal Rabi splittings, the formation of Mollow triplets and Autler-Townes doublets, ac Zeeman shifts, and even nonlinear multiphoton coherences.
\end{abstract}

\vspace{0.0cm}
\maketitle

Detection of intrinsic and random spin fluctuations -- ``spin noise" -- has attracted considerable interest in recent years, spurred by significant advances in ultrasensitive magnetic measurement techniques including nitrogen-vacancy magnetometry \cite{Mamin2013, Staudacher2013}, force-detected magnetic resonance \cite{Degen2007, Poggio2009, Poggio2013}, optical detection \cite{Sorensen98, Crooker04, Shah2010, Kuhlmann2013}, and conventional nuclear magnetic resonance \cite{MullerPNAS2006, ChandraJPCL2013}. These advances have enabled studies of ever-shrinking volumes (and numbers $n$) of spins, wherein the magnitude of statistical fluctuations ($\sim$$\sqrt{n}$) becomes increasingly important as $n$ decreases and can even exceed the average thermal spin polarization due to an applied field  -- a fact exploited in recent demonstrations of nanometer-scale nuclear spin imaging \cite{Mamin2013, Staudacher2013, Degen2009}.

For electron spins, the method of `optical spin noise spectroscopy' has emerged as a powerful technique for detecting spin dynamics that is based on passively measuring (typically via optical Faraday rotation) the stochastic spin fluctuations of an unperturbed system in strict thermal equilibrium \cite{Crooker04, ZapasskiiR, OestreichPhysE}. Initially demonstrated in atomic vapors \cite{AZ81, Crooker04}, it has also been applied to spins in bulk and low-dimensional semiconductors \cite{Oestreich05, Crooker09, Li12, Poltavtsev14}. In general however, the information obtained in thermal equilibrium (spin relaxation rates, \emph{g}-factors, etc) is necessarily linked to -- and constrained by -- the system's linear response functions, as mandated by the fluctuation-dissipation theorem \cite{Kubo}. While very useful for revealing the relevant energy levels and associated spin coherences available to a spin system, equilibrium spin noise spectra alone cannot in general reveal the underlying \emph{couplings} and \emph{correlations} between these coherences, nor the system's (usually very interesting) response to resonant driving fields. For these reasons, detection of spin fluctuations under \emph{nonequilibrium} conditions has long been desirable from both theoretical and experimental viewpoints \cite{Ivchenko, Nik, Glazov}.

Here we demonstrate that, in conjunction with a weak ac magnetic field $B_{\textrm{ac}}$, spin fluctuations in strictly unpolarized spin ensembles \emph{can} reveal patterns of correlation and coupling between a spin system's various energy levels, revealing coherent effects beyond thermal equilibrium and linear response. To introduce and benchmark this capability we apply it to a spin system with a non-trivial but very well understood magnetic ground state: a classical vapor of alkali atoms. Specifically, we study $^{41}$K atoms, in which the single spin-1/2 valence electron in the $4S$ state has a multi-level magnetic ground state due to hyperfine coupling with the $I=3/2$ nuclear spin. In the presence of $B_{\textrm{ac}}$, the intrinsic stochastic spin fluctuations alone will clearly and directly reveal the Rabi splittings of driven spin levels, as well as the associated formation of Mollow triplets, Autler-Townes doublets of neighboring spin coherences that share common Zeeman sublevels, the ac Zeeman effect, and even non-linear (multiphoton) coherence effects. 

Figure 1(a) shows a schematic of the spin noise experiment. A 10~mm thick glass cell contains isotopically enriched $^{41}$K metal and 10~Torr of nitrogen buffer gas.  Heating to 185~$^\circ$C gives a classical vapor of $^{41}$K atoms with density $\sim 7\times 10^{13}$/cm$^3$. The linearly-polarized output from a continuous-wave Ti:sapphire laser is detuned far ($\sim$40~GHz) from the D1 optical transition ($4S_{1/2}-4P_{1/2}$; 770.1~nm) and is weakly focused through the cell.  This detuning is much larger than any Doppler or pressure broadening of the D1 absorption linewidth ($<$10~GHz), which ensures that the laser does not pump or excite the $^{41}$K atoms to leading order.  The intrinsic and random spin fluctuations of the $4S$ valence electrons -- $\delta S_z(t)$ -- can nonetheless be detected via the optical Faraday rotation (FR) fluctuations $\delta \theta_F (t)$ that they impart on the detuned probe laser. This detection scheme is possible because of the optical selection rules in alkali atoms, and because FR depends not on absorption but rather on the right- and left-circularly polarized indices of refraction of the $^{41}$K vapor ($\theta_F \propto n^R - n^L$), which decay slowly with laser detuning \cite{Happer}. In this regard the measurement can be viewed as a non-perturbing probe of the vapor's intrinsic spin fluctuations \cite{Crooker04, Kuzmich, Katsoprinakis07, Shah2010}. $\delta \theta_F (t)$ is detected using balanced photodiodes, and the frequency power spectrum of this spin noise -- equivalent to the Fourier transform of the spin-spin correlator $\langle S_z(0)S_z(t) \rangle$ -- is computed in real time.

A small static magnetic field $B_{\textrm{dc}}$ of 5-15~G is applied along the transverse ($\hat{y}$) direction, and a weak radio-frequency (RF) magnetic field $|B_{\textrm{ac}}|\textrm{sin}(\omega_{\textrm{ac}}t)$ can be applied along $\hat{x}$. Since no fields are applied along the measurement direction ($\hat{z}$) and because no optical pumping occurs, the spin ensemble remains unpolarized and $\langle \delta S_z(t) \rangle$=0 throughout the experiment (all Zeeman sublevels are equally populated, in contrast to conventional methods for spin resonance).

\begin{figure}[tbp]
\includegraphics[width=0.49\textwidth]{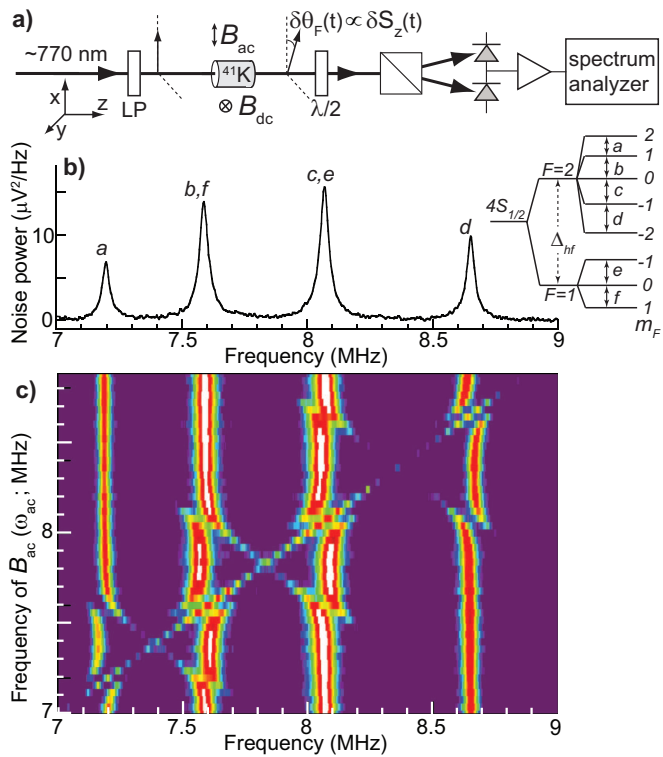}
\caption{(a) Experimental schematic. Random spin fluctuations $\delta S_z(t)$ in $^{41}$K vapor impart Faraday rotation fluctuations $\delta \theta_F(t)$ on the probe laser, which are detected by balanced photodiodes. The probe laser is detuned by $\sim$40 GHz from the $^{41}$K D1 transition ($\lambda$=770.1 nm). A static transverse magnetic field $B_{\textrm{dc}}$ is applied along $\hat{y}$, and a tunable radio-frequency field $|B_{\textrm{ac}}|\textrm{sin}(\omega_{\textrm{ac}}t)$ can be applied along $\hat{x}$. (b) Equilibrium spin noise spectrum from $^{41}$K ($T$=185~$^\circ$C, $B_{\textrm{dc}}$$\approx$11~G, $|B_{\textrm{ac}}|$=0). Noise peaks arise from spontaneous (fluctuation-induced) spin coherences between neighboring $\Delta F$=0, $\Delta m_F$=$\pm1$ Zeeman sublevels in the $^{41}$K ground state (see inset). (c) A plot of the $^{41}$K spin noise spectra as $\omega_{\textrm{ac}}$ is swept through the Zeeman coherences ($|B_{\textrm{ac}}|$$\sim$0.1~G. The induced Rabi splittings, Mollow triplets, and Autler-Townes doublets are all revealed by the intrinsic spin noise. $\langle S_z(t) \rangle$=0 throughout the measurement.} \label{Setup}
\end{figure}

Figure 1(b) shows a typical spin noise power spectrum from $^{41}$K in a transverse magnetic field ($B_{\textrm{dc}}$$\sim$11~G), under conditions of strict thermal equilibrium ($|B_{\textrm{ac}}|$=0). As observed previously \cite{Mihaila06}, the spin noise spectrum in this frequency range consists of four discrete peaks. These noise peaks are due to random spin fluctuations $\delta S_z$ of the $^{41}$K valence electrons, that are forced to precess about $B_{\textrm{dc}}$ at MHz frequencies. Equivalently, these noise peaks can be regarded as spontaneous (fluctuation-induced) spin coherences between adjacent Zeeman sublevels of the $^{41}$K ground state. Due to hyperfine coupling between the $I$=3/2 nuclear spin and the spin-1/2 valence electron, the ground state of $^{41}$K consists of two hyperfine manifolds with total spin $F$=2 and $F$=1, separated by $\Delta_{hf}$=254~MHz, as depicted in the inset. These spin manifolds split in $B_{\textrm{dc}}$ into Zeeman sublevels characterized by spin projection $m_F$. These sublevels are not equally spaced in energy due to hyperfine coupling, hence the six allowed $\Delta F$=0, $\Delta m_F$=$\pm$1 spin coherences (magnetic dipole transitions) appear at slightly different frequencies, as observed and as labeled ($a$-$f$) on the diagram and on the spin noise spectrum. Transitions $b$ and $c$ in the upper $F$=2 spin manifold are nearly degenerate with transitions $f$ and $e$ in the $F$=1 manifold, respectively, and are not resolved here.

Although very informative, equilibrium spin noise spectra alone do not uniquely reveal the detailed structure of the magnetic ground state. For example, a noise spectrum similar to that in Fig. 1(b) could in principle arise from four independent spin-1/2 species with slightly different \emph{g}-factors. In particular, an equilibrium noise spectrum does not tell us whether or how the observed spin coherences are correlated with one another (\emph{i.e.} whether they share common Zeeman sublevels), or how they couple to resonant driving fields.

These couplings and correlations can, however, be directly revealed by spin noise studies combined with the additional weak RF magnetic field $B_{\textrm{ac}}$. Figure 1(c) shows an intensity plot of the measured $^{41}$K spin noise power spectra ($x$-axis) as $\omega_{\textrm{ac}}$ is swept through the spin manifolds ($y$-axis). Clearly the intrinsic spin fluctuations are significantly influenced by $B_{\textrm{ac}}$, and several noteworthy features are immediately apparent that appear to lowest (linear) order in the field amplitude $|B_{\textrm{ac}}|$:

i) The spin noise peaks (noise coherences) exhibit a splitting when resonant with $B_{\textrm{ac}}$, as clearly seen when driving the noise peaks $a$ and $d$ (at 7.2 and 8.65~MHz). As shown in detail below, these driven noise coherences split into Mollow-type triplets -- a clear signature of coherent atom-field coupling of a two-level system to a driving field and its associated Rabi splitting of the levels.

ii) Neighboring noise coherences \emph{also} exhibit a splitting -- an Autler-Townes doublet -- demonstrating that they are coupled to the driven transition and share a common Zeeman sublevel. For example, noise peak $b$ (at 7.6 MHz) is split when $a$ is driven resonantly at 7.2 MHz.  Moreover, spin coherences $a$ and $c$ are split (but not $d$) when $b$ is driven at 7.6 MHz. This noise splitting directly reveals and quantifies the Rabi splitting of the shared level.

iii) Degenerate spin coherences such as $b$ and $f$ are generally split by different amounts by $B_{\textrm{ac}}$, thereby revealing their degeneracy and quantifying their different coupling strengths, as shown below.

\begin{figure}[tbp]
\includegraphics[width=0.48\textwidth]{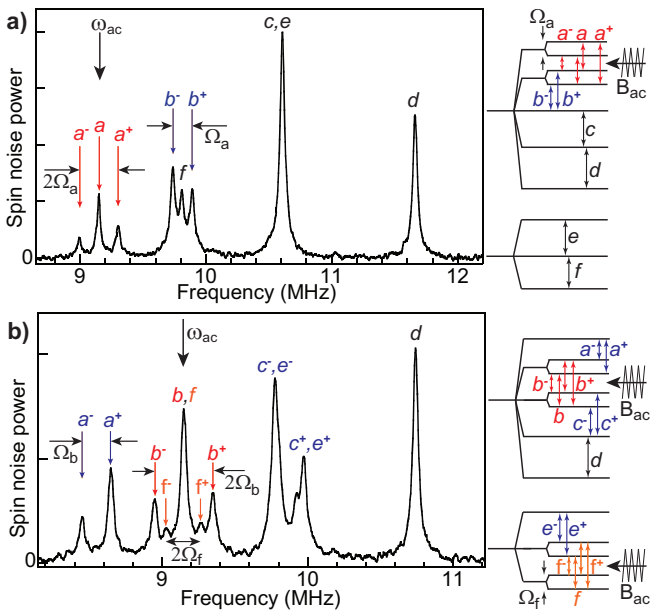}
\caption{(a) $^{41}$K spin noise spectrum when $B_{\textrm{ac}}$ is resonant with the lowest-energy Zeeman coherence $a$ ($\omega_{\textrm{ac}}=\omega_a$). (b) Noise spectrum when $B_{\textrm{ac}}$ resonant with the degenerate coherences $b$ and $f$. Diagrams show the Rabi splittings of the relevant Zeeman sublevels, and all allowed transitions, which show up clearly in the spin noise. Red and blue arrows indicate the Mollow triplet and the Autler-Townes doublet transitions respectively; black arrows indicate unperturbed coherences. (Note that for these spectra $\omega_{\textrm{ac}}$ was fixed; specific coherences were tuned into resonance via $B_{\textrm{dc}}$.)}
\label{spdr}
\end{figure}

Figure 2 shows detailed spin noise spectra corresponding to these scenarios. Here, and as discussed later, it is convenient to adopt the picture of (Zeeman) levels ``dressed" by photons of the (RF) driving field \cite{Tannoudji92, Novikov78}.  Exactly on resonance, upper and lower Zeeman sublevels of a driven magnetic dipole transition $i$ are coupled by $B_{\textrm{ac}}$ and are each split by the Rabi energy $\hbar\Omega_i= |\mu_i B_{\textrm{ac}}|$ into two states that are equal admixtures of the unperturbed sublevels. The diagrams depict the relevant Rabi splittings, and all allowed spin coherences in this frequency range.

In Fig. 2(a), the lowest-energy (and nondegenerate) spin coherence $a$ is resonant with $B_{\textrm{ac}}$. The associated sublevels $|F, m_F\rangle$=$|2,2\rangle$ and $|2,1\rangle$ are each split by $\Omega_a$, leading to \emph{four} possible spin coherences as depicted by the red arrows in the diagram. The middle two are necessarily degenerate, giving three transitions that are observable in the spin noise spectrum: This is the spin/magnetic analog of the well-known Mollow triplet that appears when driving electric dipole transitions with resonant light (\emph{i.e.}, ac electric fields) \cite{Mollow69, Wu75, Xu2007, Flagg2009}.  It is characterized by unshifted central peak $a$ and satellites $a^+$ and $a^-$ separated by $2\Omega_a$.

Moreover, it is clear that the neighboring spin noise coherence $b$ is also split into an Autler-Townes doublet \cite{Autler50, Xu2007}, indicating that it shares a common Zeeman sublevel with the driven transition (namely, $|2,1\rangle$). The diagram depicts the new transitions $b^+$ and $b^-$, which are split by $\Omega_a$. Note that the splitting of coherence $b$ reveals the otherwise degenerate spin coherence $f$ in the lower $F$=1 manifold, which shares no common sublevels with transition $a$ and is therefore unaffected by $B_{\textrm{ac}}$. Coherences $c$, $d$, and $e$ also share no common sublevels with $a$ and are similarly unperturbed.

A slightly more complex scenario is shown in Fig. 2(b), where the nearly-degenerate transitions $b$ and $f$ are resonant with $B_{\textrm{ac}}$. Here, a pair of Zeeman sublevels in \emph{both} the $F$=2 and $F$=1 manifold show a Rabi splitting ($\Omega_b$ and $\Omega_f$). That two different transitions are driven by $B_{\textrm{ac}}$ is easily seen by the fact that the spin noise peak splits into \emph{five} distinct peaks (instead of three), arising from two overlapping Mollow triplets with different Rabi frequency. Note also the Autler-Townes doublet of the neighboring coherence \emph{a}, and the two overlapping Autler-Townes doublets from transitions \emph{c} and \emph{e}. As expected, the spin coherence \emph{d} is unaffected.

\begin{figure}[tbp]
\includegraphics[width=0.42\textwidth]{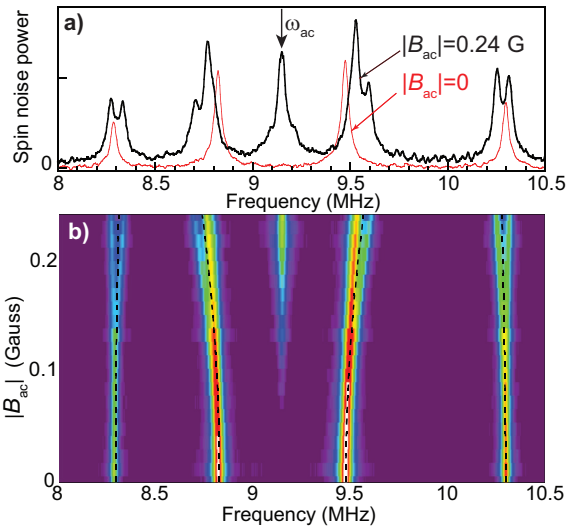}
\caption{$^{41}$K spin noise spectra in the presence of a nonresonant $B_{\textrm{ac}}$. $\omega_{\textrm{ac}} =\frac{1}{2}(\omega_b + \omega_c)$, $B_{\textrm{dc}}\approx 13$ G. AC Zeeman shifts ($\propto |B_{\textrm{ac}}|^2$) of all noise coherences are evident. Dotted lines show expected ac Zeeman shifts computed to lowest order.  Additional splittings of the spin noise peaks emerge at large $|B_{\textrm{ac}}|$, due to higher-order multi-photon coupling (see text).} \label{rfcentered}
\end{figure}

As these experiments demonstrate, optical studies of stochastic spin fluctuations provide a convenient route to simultaneously detect the allowed transitions and coherences within the magnetic ground state, and to directly observe how they are coupled to each other and to external fields by monitoring their response to an RF perturbation. The patterns of correlations between noise peaks and the multiplicity of splittings can be used to reconstruct the detailed structure of the magnetic ground state, and to study coherent effects beyond thermal equilibrium and linear response.  Note that this information is revealed via the stochastic spin fluctuations alone, using a strictly unpolarized atomic vapor ($\langle S_z(t) \rangle$=0): at no point is the $^{41}$K spin ensemble optically polarized or pumped in any way (in contrast to conventional methods for atomic spin resonance or RF double-resonance \cite{Brossel52, Steinfeld78}, which generally require optical pumping and a nonequilibrium ensemble polarization). Moreover, passive spin detection via the Faraday effect using far-detuned probe lasers ensures insensitivity to Doppler broadening effects and correspondingly well-resolved spin transitions.  In condensed matter systems such as interacting quantum dots or in atomically-thin materials, these noise techniques may prove highly desirable, since conventional methods typically rely on direct optical pumping of spin polarization and the associated formation of excitons and unavoidable heating.

Additionally, off-resonant perturbations can also be clearly revealed via spin fluctuations. Figure 3 shows the evolution of the noise spectra as $B_{\textrm{ac}}$, which is detuned midway between spin coherences $b$ and $c$ [$\omega_{\textrm{ac}} = (\omega_b + \omega_c)/2$], is increased in amplitude. At low amplitude the spin noise spectra are unperturbed, as expected. With increasing RF amplitude all noise coherences visibly shift.  (Further, they exhibit a splitting at the largest $|B_{\textrm{ac}}|$, discussed below.) The shift is a direct manifestation of the ac Zeeman effect -- \emph{i.e.}, the spin/magnetic analog of the ac Stark shift that is well known in atomic and semiconductor physics \cite{Tannoudji92, Budker} when an optical transition is excited by an off-resonant laser. Within the rotating-wave approximation, the shifts are quadratic in $|B_{\textrm{ac}}|$, and upper/lower Zeeman sublevels of transition $i$ shift by an amount $\Delta E /\hbar = \pm \Omega_i^2 /4(\omega_i - \omega_{\textrm{ac}})$ \cite{Tannoudji92, Budker, Ramsey, CT}.

\begin{figure}[tbp]
\includegraphics[width=0.45\textwidth]{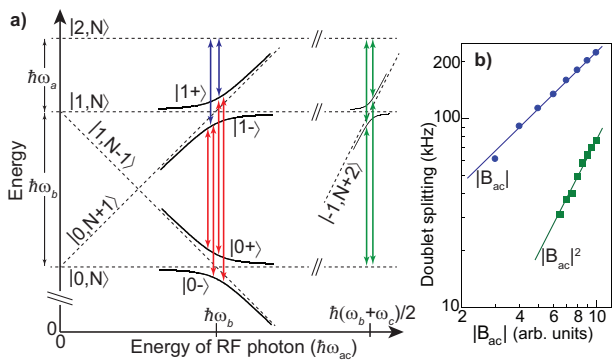}
\caption{(a) Diagram showing the upper three Zeeman sublevels ($m_F$=2,1,0) of the $^{41}$K $F$=2 manifold, in the ``dressed state" picture, versus RF photon energy $\hbar\omega_{\textrm{ac}}$. Uncoupled dressed states are labeled by $|m_F$, \#photons$\rangle$ and appear as dashed straight lines. Solid curved lines show the new coupled states that anticross. Mollow triplet (red arrows) and Autler-Townes doublet transitions (blue) are indicated. A higher-order (two-photon) anticrossing is shown at the right. (b) The measured doublet splitting is linear (quadratic) in $|B_{\textrm{ac}}|$ for the case of one- (two-) photon anticrossing.} \label{SN1}
\end{figure}

All these spin noise data can be formally understood and readily visualized within a picture of atomic states that are ``dressed" by photons of a driving field (\emph{i.e.}, Zeeman sublevels dressed by RF photons of $B_{\textrm{ac}}$) \cite{Tannoudji92, Novikov78}. Figure 4(a) captures the essential physics by showing a few of the dressed Zeeman sublevels versus the RF photon energy $\hbar \omega_{\textrm{ac}}$. Uncoupled dressed states, labeled as $|m_F$, \#photons$\rangle$, appear as straight dashed lines. Thus, $|0,N\rangle$ is an atom in the $m_F$=0 state in the presence of $N$ photons ($N$ is the mean photon number to generate $|B_{\textrm{ac}}|$), while $|m_F,N \pm 1 \rangle$ is an atom in the presence of one additional (or one less) photon; these are diagonal lines with slope $\pm$1.

Intersecting dashed lines coupled by the atom-field interactions (magnetic dipole coupling $|\mu_i B_{\textrm{ac}}| = \hbar\Omega_i$), lead to an anticrossing and new hybrid levels (solid curved lines).  For example, exactly on resonance with transition $b$ (when $\omega_{\textrm{ac}}=\omega_b$), the new coupled states $|1+\rangle$ and $|1-\rangle$ are equal (and orthogonal) superpositions of the unperturbed dressed states $|1,N \rangle$ and $|0,N+1 \rangle$, separated by the bare Rabi energy $\hbar \Omega_b = |\mu_b B_{\textrm{ac}}|$. Since both $|1+\rangle$ and $|1-\rangle$ `contain' $|1,N \rangle$, two transitions to the neighboring level $|2,N \rangle$ are allowed (blue arrows) -- this is the Autler-Townes doublet observed, \emph{e.g.}, in noise coherence $a$ in Fig 2(b). Similarly, four transitions (red arrows) exist between the coupled states $|1\pm \rangle$ and $|0 \pm \rangle$.  Two are necessarily degenerate, giving the Mollow-type triplet observed in the spin noise spectrum. The complex evolution of the magnetic ground state as $\omega_{\textrm{ac}}$ is tuned through the spin manifolds can therefore be understood in a straightforward manner within the dressed-state picture, and directly compared with the measured spin noise as shown in Fig. 1(c) and Fig 2.

In the dressed-state picture, the additional splittings of the coherences observed in Fig. 3 for large $|B_{\textrm{ac}}|$ can immediately be recognized as Autler-Townes doublets (green arrows in Fig. 4) caused by \emph{two-photon} anticrossings between levels $|1,N \rangle$ and $|-1,N+2 \rangle$ (and between $|-1,N \rangle$ and $|1,N-2 \rangle$).  This occurs when $\omega_{\textrm{ac}}=(\omega_b + \omega_c )/2$, as in Fig. 3. These two-photon splittings are verified to scale as $|B_{\textrm{ac}}|^2$, rather than as $|B_{\textrm{ac}}|$ for the ordinary Autler-Townes doublet (Fig. 4b). Thus, although the spin ensemble remains unpolarized, the system's intrinsic spin fluctuations can nonetheless reveal even higher-order and nonlinear coupling to external perturbations.

In summary, optical spin noise spectroscopy in non-equilibrium conditions is shown to reveal the full spectrum of allowed and available spin coherences, and their coherent coupling to external perturbations. In this demonstration using $^{41}$K vapor, couplings and correlations between Zeeman sublevels are plainly revealed by the intrinsic spin noise, from which the detailed structure of the magnetic ground state can be inferred. Coherent phenomena including Rabi splittings, magnetic Mollow triplets, Autler-Townes doublets, and ac Zeeman shifts are clearly shown via the spin noise, in addition to higher-order (multi-photon) coupling. Importantly and more generally, this information is accessible from an unpolarized spin ensemble ($\langle S_z(t) \rangle$=0), providing a minimally-perturbative and fluctuation-based alternative route to access coherent phenomena in spin systems beyond standard linear response.

We gratefully acknowledge helpful discussions with D. L. Smith, V. S. Zapasskii and E. B. Aleksandrov.  This work was supported by the Los Alamos LDRD program and the Deutsche Forschungsgemeinschaft.


\end{document}